\documentclass[prl,twocolumn,showpacs,showkeys]{revtex4}

\begin{document}

\title{Translation invariant topological superconductors on lattice}

\author{Su-Peng Kou}
\affiliation{Department of Physics, Beijing Normal University,
Beijing, 100875 P. R. China }

\author{Xiao-Gang Wen}
\homepage{http://dao.mit.edu/~wen}
\affiliation{Department of
Physics, Massachusetts Institute of Technology, Cambridge,
Massachusetts 02139}

\begin{abstract}
In this paper we introduce four $Z_2$ topological indices $\zeta_{\v
k}=0,1$ at $\v {k}=(0,0)$, $(0,\pi )$, $(\pi ,0)$, $(\pi ,\pi )$
characterizing 16 universal classes of 2D superconducting states
that have translation symmetry but may break any other symmetries.
The 16 classes of superconducting states are distinguished by their
even/odd numbers of fermions on even-by-even, even-by-odd,
odd-by-even, and odd-by-odd lattices.  As a result, the 16 classes
topological superconducting states exist even for interacting systems.
For non-interacting systems, we find that $\zeta_{\v k}$ is the number
of electrons on $\v {k}=(0,0)$, $(0,\pi )$, $(\pi ,0)$, or $(\pi
,\pi)$ orbitals (mod 2) in the ground state.  For 3D superconducting
states with only translation symmetry, topological indices 
give rise to 256 different types of topological superconductors.
\end{abstract}


\maketitle

\emph{Introduction}:
In last 20 years, it became more and more clear that Landau symmetry
breaking theory \cite{L3726,GL5064,LanL58} cannot describe all possible
orders in quantum states of matter (the states of matter at zero
temperature).\cite{Wrig} The new order is called topological
order for gapped states.  Fractional quantum hall
systems\cite{TSG8259} and many other systems were shown to have
topologically ordered ground
states.\cite{KL8795,WWZcsp,RS9173,Wsrvb,Wtoprev,K032,MS0181,BFG0212,MSP0202,Wqoexct,FK0997}

Topological order can exist even if we break all symmetries.  However,
for systems with certain symmetries, a new type of order - symmetry
protected topological order - can appear.\cite{Wqoslpub,GW0969}  Even
though the ground states with symmetry protected topological order do
not break any symmetries, they can still represent different phases of
matter.  The simplest example of symmetry protected topological orders
is the $Z_2$ topological insulator that can appear in 2D and 3D free
fermion systems with time reversal
symmetry.\cite{MNZ0404,KM0501,KM0502,MB0706,FKM0703,QHZ0837} The
Haldane phase in 1D spin-1 chain is the oldest example of symmetry
protected topological phase with time reversal, parity and translation
symmetries.\cite{H8353,AKL8877,GW0969} Symmetry protected topological
order appear quite commonly in topological phases with symmetries.
The projective symmetry group is introduced to (partially)
characterize/distinguish different symmetry protected topological
orders.\cite{Wqoslpub}

In this paper, we study 2D fully gapped superconducting (SC) states on
a lattice that have only translation symmetry.  The time reversal,
spin rotation, lattice 180$^\circ$ rotation and parity \etc may not be
the symmetries of the SC Hamiltonian.  The time-reversal violating SC
states can be characterized by a winding munber.\cite{SMF9945,RG0067} We found
that the 2D SC states with a given winding number can be
further divided into 8 classes.  Although all those classes of SC states
have the same symmetry, they cannot change into each other without
quantum phase transitions which close the energy gap. So the different
classes of SC states belong to different quantum phases are called
topological superconductors.  Our results can be easily generalized to
3D lattice which leads to 256 different topological superconductors
with translation symmetry.

If we consider different symmetries (other than the translation
symmetry), then different classes of topological superconductors can
be obtained.  In particular, the time-revisal invariant topological
superconductors are studied in \Ref{R0664,QHR0901,SF0904}.  The non-tranlation
invariant topological superconductors with different symmetry classes
are studied in \Ref{SRF0825,K0986}.

In 2D, four $Z_2$ topological indices $\zeta_{\v k}$ at $\v
{k}=(0,0)$, $(0,\pi )$, $(\pi ,0)$, $(\pi ,\pi )$ are introduced to
characterize 16 classes of topological SC states.  The 16
classes of SC states are distinguished by their even/odd
number of fermions on even-by-even, even-by-odd, odd-by-even, and
odd-by-odd lattices.  We stress that the topological SC
states discussed here exist even for interacting systems.  Also, the
topological indices $\zeta_{\v k}$ can distinguish different symmetry
protected topological orders that cannot be distinguished by
projective symmetry group.

\emph{Fully gapped 2D SC states with only translation symmetry}:
We use $\v i=(i_x,i_y)$ to label unit cells of a lattice and
$\al=1,2,...$ to label the electron operators $\psi_{\al,\v i}$ in the
unit cell $\v i$.  The index $\al$ labels the spin and/or orbitals of
electrons.  The 16 classes of topological SC states can already exist
when $\al=1$ (\ie only one electronic state per unit cell).  So
without losing generality, we will assume $\al=1$ here and drop the
$\al$ index.  In this case, the most general SC state with only
translation symmetry can be described by
\begin{equation}
H=\sum_{\v i\v j}\psi _{\v i}^{\dag }u_{\v i\v j}\psi _{\v j}
+\sum_{\v i\v j}(\psi _{\v i}^{\dag }\eta _{\v i\v j}\psi _{\v j}^{\dag }
+h.c.),
\label{mean}
\end{equation}
where $u_{\v i\v j}$ and $\eta _{\v i\v j}$ and complex numbers.  The
translation invariance  requires that $u_{\v i\v j}=u_{\v i+\v a,\v j+\v a}$ and
$\eta_{\v i\v j}=\eta_{\v i+\v a,\v j+\v a}$.

One can rewrite the SC Hamiltonian in momentum space by introducing
$\Psi _{\v {k}}=\left(
\begin{array}{l}
\psi _{\v {k}} \\
\psi _{-\v {k}}^{\dag } \\
\end{array}
\right)$
and $ \Psi _{\v {k}}^{\dag }=\left(
\begin{array}{ll}
\psi _{\v {k}}^{\dag } & \psi _{-\v {k}}
\end{array}
\right) $.
Note that $\Psi _{\v {k}}$ satisfy the following algebra
\[
\{\Psi _{I\v {k}}^{\dag },\Psi _{J\v {k}^{\prime }}\}=\delta
_{IJ}\delta _{\v {k}-\v {k}^{\prime }},\ \ \ \  \{\Psi _{I\v {
k}},\Psi _{J\v {k}^{\prime }}\}=(\si_1)_{IJ}\delta _{\v {k}+\v {k}
^{\prime }},
\]
where $\si_l$, $l=1,2,3$ are Pauli matrices.
We also note that $(\Psi _{-\v {k}
}^{\dag },\Psi _{-\v {k}})$ can be expressed in term of $(\Psi _{\v {k}
}^{\dag },\Psi _{\v {k}})$:
\begin{equation}
\Psi _{-\v {k}}=\si_1 \Psi _{\v {k}}^{*},\ \ \ \ \ \ \ \ \Psi _{-
\v {k}}^{\dag }=\Psi _{\v {k}}^T\si_1 .
 \label{Psikmk}
\end{equation}

In terms of $\Psi _{\v {k}}$, $H$ can be written as
\begin{equation}
H=
\sum_{\v {k}\neq 0}\Psi _{\v {k}}^{\dag }M(\v {k})\Psi _{\v {k}}
+\sum_{\v {k} = 0}\Psi _{\v {k}}^{\dag }M(\v {k})\Psi _{\v {k}}
\label{Hmean1}
\end{equation}
where $-\pi < k_x,k_y <+\pi$ and $M(\v {k})$ are $2\times 2$
Hermitian matrices $M(\v {k})=M^\dag(\v {k})$.  Here $\v {k}=0$ means
that $(k_x,k_y)=(0,0)$, $(0,\pi )$, $(\pi ,0)$, or $(\pi ,\pi )$.
Also $k_x$ and $k_y$ are quantized: $k_x=\frac{2\pi}{L_x}\times$
integer and $k_y=\frac{2\pi}{L_y}\times$ integer, where $L_x$ and
$L_y$ are size of the square lattice in the $x$- and $y$-directions.
In paper, we assume the periodic boundary condition.

Note that on an even by even lattice (\ie
$L_x=$ even and $L_y=$ even), $(k_x,k_y)=(0,0)$, $(0,\pi )$, $(\pi
,0)$, or $(\pi ,\pi )$ all satisfy the quantization conditions
$k_x=\frac{2\pi}{L_x}\times$ integer and $k_y=\frac{2\pi}{L_y}\times$
integer.  In this case, $\sum_{\v {k} = 0}$ sums over all the four
points $(k_x,k_y)=(0,0)$, $(0,\pi )$, $(\pi ,0)$, and $(\pi ,\pi )$.
On other lattices, $\sum_{\v {k} = 0}$ sums over less points.  Say on
an odd by odd lattice, only $(k_x,k_y)=(0,0)$ satisfies the
quantization conditions $k_x=\frac{2\pi}{L_x}\times$ integer and
$k_y=\frac{2\pi}{L_y}\times$ integer.  In this case,  $\sum_{\v {k} =
0}$ sums over only $(k_x,k_y)=(0,0)$ point.

We note that
$\Psi _{\v {k}}^{\dag }\Psi _{\v {k}}=2$.
Thus up to a constant in $H$, we may assume $M(\v {k})$
to satisfy $\mathrm{Tr}M(\v {k})=0$.
Due to Eq. (\ref{Psikmk}),
\[
\Psi _{-\v {k}}^{\dag }M(-\v {k})\Psi _{-\v {k}}=\mathrm{Tr}M(
\v {k})-\Psi _{\v {k}}^{\dag }\si_1 M^T(-\v {k})\si_1 \Psi _{
\v {k}}.
\]
Thus, we may rewrite Eq. (\ref{Hmean1}) as
\begin{align}
\label{meanU}
H &=
\sum_{\v {k}>0}\Psi _{\v {k}}^{\dag }U( \v {k})\Psi _{\v {k}}
+\frac12 \sum_{\v {k}=0}\Psi _{\v {k}}^{\dag }U( \v {k})\Psi _{\v {k}}
,
\nonumber\\
U(\v k)&=M(\v k) -\si_1 M^T(-\v k) \si_1 .
\end{align}
Here
$\v {k}>0$ means that $\v {k}\neq 0$ and $k_y>0$ or $k_y=0,\ k_x>0$.
Clearly $U( \v {k})$ satisfies
\begin{equation}
\label{UU}
U(\v {k})=-\si_1 U^T(\v {-k})\si_1 ,\ \ \
U(\v {k})=U^\dag(\v {k}) .
\end{equation}

Now we expand the traceless $U(\v {k})$ by $3$ Pauli matrices
$\si_l$. We have
\[
U(\v {k})=\sum_{\{\alpha ,\beta \}}c_l(\v {k}
)\si_l
\]
where $c_l(\v {k})$ are real.
From \eq{UU}, we find $c_3(\v {k})=c_3(-\v {k})$ and $c_l(\v
{k})=-c_l(-\v {k})$, $l=1,2$.  Thus for odd matrices, $c_l(\v {k})$
are zero at momentum $(0,0)$, $(0,\pi )$, $(\pi ,0)$, $(\pi ,\pi )$.

\emph{Classification of topological superconductors}:
For a generic choice of $u_{\v i\v j}$ and $\eta_{\v i\v j}$, the corresponding SC
Hamiltonian \eq{meanU} is gapped.  Note that the energy levels of the
SC Hamiltonian \eq{meanU} appear in $(E,-E)$ pairs. The SC ground
state is obtained by filling all the negative energy levels.  The SC
Hamiltonian is gapped if the minimal positive energy is finite.

As we change the SC ansatz $u_{\v i\v j}$ and $\eta_{\v i\v j}$, the
SC energy gap may close which indicate a quantum phase transition.
Thus if two gapped regions are always separated by a gapless region,
then the two gapped regions will correspond to two different phases.
We may say that the two phases carry different topological orders.

In the following, we introduce topological indices that can be
calculated for each gapped SC ansatz $(u_{\v i\v j},\eta_{\v i\v j})$.  We
will show that two gapped SC ansatz with different topological
indices cannot smoothly deform into each other without closing the
energy gap.  Therefore, the topological indices characterize different
SC states translation symmetry.

The SC Hamiltonian in momentum space \eq{meanU}, has a form
$H=H(\v{k}>0) +H( \v{k}=\v{0})$.
First, let us diagonalizing the SC Hamiltonian at the points
$\v{k}>0$. Introducing
\begin{align}
W(\v k) \Psi _{\v{k}}&=\left(
\begin{array}{l}
\alpha _{\v{k}} \\
\alpha _{-\v{k}}^{\dag } \\
\end{array}
\right)
,
\end{align}
where
\begin{align}
 W(\v k)U(\v{k})W^\dag(\v k)
=\left(
\begin{array}{llll}
\varepsilon (\v{k}) & 0 \\
0 & -\varepsilon (\v{k})\\
\end{array}
\right) ,\ \ \
\varepsilon(\v{k})>0,
\end{align}
we find
\begin{eqnarray*}
H(\v{k} >0)=\sum_{\v{k}>0}
\varepsilon(\v{k})
(\alpha _{\v k}^{\dag }\alpha _{\v k}- \alpha_{-\v k}\alpha_{-\v k}^\dag)
\end{eqnarray*}
We note that $\alpha _{\pm \v{k}}$
will annihilate the SC ground state:
\[
\alpha _{\pm \v{k}}|\Psi _{\text{SC}}\rangle =0 .
\]
At the four $\v k=0$ points, the Hamiltonian is already diagonal
since $c_{1,2}(\v k)=0$.

The energy spectrum at $\v k=0$ motivates
us to introduce four $\eta(\v k)$ as
the topological indices, one for each $\v k=0$ point:
\begin{align}
\label{etac}
\zeta_{\v k} &=1-\Th [c_3(\v k)]
\end{align}
where $\Th(x)=1$ if $x>0$ and $\Th(x)=0$ if $x<0$.  If two SC states
have different sets of topological indices $(\zeta_{\v k=(0,0)},
\zeta_{\v k=(\pi,0)}, \zeta_{\v k=(0,\pi)}, \zeta_{\v k=(\pi,\pi)})$,
then as we deform one state smoothly into the other, some $\zeta_{\v
k}$ must change sign.  When  $\zeta_{\v k}$ change sign, then $c_3(\v
k)=0$ and the SC state becomes gapless indicating a quantum phase
transition.  Therefore, there are 16 different translation invariant
SC labeled by $\zeta_{\v k=(0,0)}$, $\zeta_{\v k=(0,\pi )}$, $\eta _{\v
k=(\pi,0)}$, $\eta _{\v k=(\pi ,\pi )} =$ $\mathrm{1111}$,\textrm{\
}$\mathrm{1100}$,\textrm{\ }$\mathrm{1010 }$,\textrm{\
}$\mathrm{1001}$,\textrm{\ }$\mathrm{0101}$,\textrm{\ 0011}, \textrm{\
0110},\textrm{\ 0000},\textrm{\ 1000},\textrm{\ 0100},\textrm{\
0010},\textrm{\ 0001},\textrm{\ 1110},\textrm{\ 1101},\textrm{\ 1011},
\textrm{\ 0111}.

\emph{The physical quantum numbers separating topological SC
states}:
In the above, we introduced 16 classes of translation invariant SC
states through the four topological indices $\zeta_{\v k=0}$.
However, as we deform one class of SC state to another, we have
assumed the range of spin/orbital index $\al$ to be $\al=1$. If the
range of $\al$ is more than $1$, do we still have to encounter gapless
region as we deform one class of SC state to another? Also, if
electrons are interacting, whether different classes of SC states
are still separated by gapless region?

In the following, we show that even with many spin/orbital states per
unit cell and even in the presence of weak interactions, there are
still 16 classes of translation invariant SC states.  We obtain this
result by finding universal physical quantum numbers that separate the 16
classes of SC states. The universal physical quantum numbers are
$(-)^{N_e}$ on even-by-even, odd-by-even, even-by-odd, and odd-by-odd
lattices.  Here $N_e$ is the number of electrons in the SC ground
state.  Note that $(-)^{N_e}$ commutes with the SC Hamiltonian.
Although $N_e$ is not definite in the SC ground state, $(-)^{N_e}$ is
uniquely defined.

We note that $N_e=N_{\v k\neq 0}+N_{\v {k}=0}$ where $N_{\v {k}\neq \v
{0}}=\sum\limits_{ \v {k>0}}\psi _{\v {k}}^{\dag }\psi _{\v
{k}}+\sum\limits_{ \v {k<0}}\psi _{\v {k}}^{\dag }\psi _{\v {k}}$ and
$N_{\v {k} =0}=\sum\limits_{\v {k}=0}\psi _{\v {k}}^{\dag }\psi _{ \v
{k}}$.
For $\v {k}>0$, we have
\begin{align}
&\ \ \
(-)^{
\psi _{\v {k}}^{\dag }\psi _{\v {k}}
+\psi _{-\v {k}}^{\dag }\psi _{-\v {k}}
}
=
-(-)^{
\psi _{\v {k}}^{\dag }\psi _{\v {k}}
-\psi _{-\v {k}}\psi_{-\v {k}}^\dag
}
\nonumber\\
&=
-(-)^{
\psi _{\v {k}}^{\dag }\psi _{\v {k}}
+\psi _{-\v {k}}\psi_{-\v {k}}^\dag
}
=
-(-)^{
\alpha _{\v {k}}^{\dag }\alpha _{\v {k}}
+\alpha _{-\v {k}}\alpha _{-\v {k }}^\dag
}
\nonumber\\
&=
(-)^{
\alpha _{\v {k}}^{\dag }\alpha _{\v {k}}
+\alpha _{-\v {k}}^\dag\alpha _{-\v {k}}
}
\end{align}
Hence we have
\begin{eqnarray}
&&(-1)^{\psi _{\v {k}}^{\dag }\psi _{\v {k}
}+\psi _{-\v {k}}^{\dag }\psi _{-\v {k}}}\mid \Psi_{\mathrm{
SC}}\rangle \>
=(-1)^{\alpha _{\v {k}}^{\dag }\alpha _{\v {k}}
+\alpha _{-\v {k}}^\dag\alpha _{-\v {k}}
}\mid \Psi _{\mathrm{SC}} \rangle
\nonumber\\
&=&\mid \Psi _\text{SC}\rangle \>
\end{eqnarray}
for $\v {k}>0$. The total number of the electrons on all the
$\v {k}>0$ orbitals is always even.

So to determine if the SC ground state contain even or odd number of
electrons, we only need to count the number of the electrons at the
$\v {k}=0$ points.  At $\v k=0$, we have
\begin{align}
\label{Nezeta}
\psi _{\v k}^\dag \psi _{\v k} |\Psi_\text{SC}\>
=(1-\Th[c_3(\v k)])|\Psi_\text{SC}\>=\zeta_{\v k}|\Psi_\text{SC}\>
\end{align}
We see that for non-interacting electrons, the
topological indices $\zeta_{\v k}$ at the $\v k=0$ points are just the
numbers of electrons in the SC ground state on the corresponding $\v k$
orbitals mod 2.  This can be used as a definition of topological
indices.  If the gapped SC phase has a weak SC order, then $\zeta_{\v
k}$ are just the numbers of electrons in the normal state on the
corresponding $\v k$ orbitals mod 2.  From the above discussion, we see
that spin-singlet SC states always have $\{\zeta_{\v k}\}=0000$
topological SC order.

From \eq{Nezeta}, we find that the total fermion number at the $\v
k=0$ points and the total number of electrons are given by
\begin{equation}
\label{Ne2}
N_{\v {k}=\v {0}} \text{ mod }2
=N_e \text{ mod } 2
=\sum\limits_{\v {k=}0} \zeta_{\v k} \text{ mod } 2
\end{equation}
Note that on even-by-even lattice (ee), all the
four $\v k=0$ points $\v {k}=(0,0)$, $(0,\pi )$, $(\pi ,0)$, $(\pi
,\pi )$ are allowed.  In this case $N_e$ mod 2 is the sum of all four
$\zeta_{\v k=0}$ mod 2.  On even-by-odd lattice (eo), only two $\v k=0$
points $\v {k}=(0,0)$, $(\pi ,0)$ are allowed.  In this case $N_e
\text{ mod } 2 = \zeta_{(0,0)}+ \zeta_{(\pi,0)}$ mod 2.
This way, \eqn{Ne2} allows us to construct the follow table:
\begin{align}
\label{Netbl}
\begin{array}{rcccc}
(-)^{N_e} &(ee) & (eo) & (oe) & (oo) \\
(0000): & + & + & + & + \\
(1111): & + & + & + & - \\
(0101): & + & + & - & + \\
(1010): & + & + & - & - \\
(0011): & + & - & + & + \\
(1100): & + & - & + & - \\
(0110): & + & - & - & + \\
(1001): & + & - & - & - \\
(0001): & - & + & + & + \\
(1110): & - & + & + & - \\
(0100): & - & + & - & + \\
(1011): & - & + & - & - \\
(0010): & - & - & + & + \\
(1101): & - & - & + & - \\
(0111): & - & - & - & + \\
(1000): & - & - & - & - \\
\end{array}
\end{align}
about the even/odd number of electrons in the SC ground states on
various lattices.

We see that all 16 classes of SC states have distinct even/odd number
of electrons on the four types of lattices.  Since $N_e$ mod 2 is
discrete, so it is a universal quantum number in a gapped phase that
is robust against perturbations of weak mixing with other spin/orbital
states and adding weak interactions.  Therefore, the 16 topological SC
phases is robust against weak spin/orbital mixing and weak
interactions.

\emph{Examples of translation invariant topological SC phases}:
Let us first consider the following $p_x+ i p_y$ SC state
\begin{eqnarray}
\label{Hmean11}
H &=&\sum_{\v i\v j}(\psi _{\v i}^{\dag }u_{\v i\v j}\psi _{\v j} +\psi _{\v i}\eta_{\v
i\v j}\psi _{\v
j} +h.c.)
\\
u_{i,i+x} &=&u_{i,i+y}= -\chi_1 ,  \ \ \
u_{i,i+x+y} = u_{i,i-x+y} =  -\chi_2 ,
\nonumber \\
\eta_{i,i+x} &=& \eta ,  \ \ \ \eta_{i,i+y}= i\eta .
\nonumber 
\end{eqnarray}
We find that
\begin{align}
c_3(\v k)&= -2\chi_1(\cos(k_x)+\cos(k_y))
\nonumber\\
&\ \ \    -2\chi_2(\cos(k_x+k_y)+\cos(k_x-k_y))
\nonumber\\
c_1(\v k)&= 2\eta\sin(k_x),\ \ \ \
c_2(\v k)= 2i\eta\sin(k_y).
\end{align}
Assume $\chi_1>0$, we find that when $\chi_2>0$ the SC state is a
$\{\zeta_{\v k}\}=1000$ topological superconductor.  When
$-\chi_1<\chi_2<0$, the SC state is a $\{\zeta_{\v k}\}=1110$
topological superconductor.  When $\chi_2<-\chi_1$, the SC state is a
$\{\zeta_{\v k}\}=0110$ topological superconductor.  

The $p_x+ i p_y$ topological superconductor is also characterized by a
winding munber $w$ which is given by $w=\text{sgn}(\eta)$ for the
$\chi_2>0$ case.\cite{RG0067} We like to point out that in general the
wind number satisfies 
\begin{align}
w \text{ mod }2= \zeta_{(0,0)} +\zeta_{(\pi,0)}
+\zeta_{(0,\pi)} +\zeta_{(\pi,\pi)} \text{ mod }2.
\end{align}
The above result implies that the topological indices
$\{\zeta_{\v k}\}$  do not provide a complete characterization of
topological order, \ie for a given set of $\{\zeta_{\v k}\}$, there
can still be different topological phases distinguished by some other
topological quantum numbers, such as the winding number.

Let us consider spin-1/2 SC states with spin-orbital coupling in more
detail.  We need to consider a more general case where there are two
spin/orbital states per unit cell.  In this case, the most general SC
state with only translation symmetry is described by
\begin{equation}
H=\sum_{\v i\v j}\psi _{\v i}^{\dag }u_{\v i\v j}\psi _{\v j}+\sum_{\v i\v j}(\psi
_{\v i}^{\dag }\eta _{\v i\v j}\psi _{\v j}^{\dag }+h.c.).
\label{meansp}
\end{equation}
where $u_{\v i\v j}$ and $\eta_{\v i\v j}$ are 2 by 2 matrices.
One can rewrite the SC Hamiltonian in momentum space by introducing
$\Psi _{\v {k}}^T=\left(
\psi _{1,\v {k}} ,
\psi _{1,-\v {k}}^{\dag } ,
\psi _{2,\v {k}} ,
\psi _{2,-\v {k}}^{\dag }
\right)$:
\begin{align}
\label{meanUsp}
H &=
\sum_{\v {k}>0}\Psi _{\v {k}}^{\dag }U( \v {k})\Psi _{\v {k}}
+\frac12 \sum_{\v {k}=0}\Psi _{\v {k}}^{\dag }U( \v {k})\Psi _{\v {k}}
.
\end{align}
where $U( \v {k})$ satisfies
$U(\v {k})=-\Ga U^T(\v {-k})\Ga$,
$U(\v {k})=U^\dag(\v {k})$, and $\Ga=\si_1\otimes \si_0$.
We can expand $U(\v {k})$ by $16$ Hermitian matrices
$ M_{\{\alpha \beta \}} \equiv
\si_\al\otimes \si_\bt$, $ \al,\bt =0,1,2,3 $,
where $\si_0=\v 1$.  We have
$U(\v {k})=\sum_{\{\alpha ,\beta \}}c_{\{\alpha \beta \}}(\v {k}
)M_{\{\alpha \beta \}}$,
where $c_{\{\alpha \beta \}}(\v {k})$ are real.  We find that at the
four $\v k=0$ points, only $c_{\{30\}}$, $c_{\{12\}}$, $c_{\{22\}}$,
$c_{\{33\}}$, $c_{\{31\}}$, and $c_{\{02\}}$ are non-zero.
The topological indices at
$\v k=0$ points are
\begin{align}
 \zeta_{\v k}&=1-\Th
 [
c_{\{30\}}^2(\v {k} )+c_{\{12\}}^2(\v {k})
+c_{\{22\}}^2(\v {k})
\nonumber\\
&\ \ \ \ \ \ \ \ \ \
- c_{\{33\}}^2(\v {k})-c_{\{31\}}^2(\v {k})
-c_{\{02\}}^2( \v {k})) 
].
\end{align}
This equation allows us to calculate $\zeta_{\v k}$ for
spin-1/2 superconductors that may break spin rotation symmetry.

\emph{Conclusion}:
Using the even/odd numbers of electrons at the four $\v k=0$ orbitals,
we find that a gapped 2D SC states with translation symmetry can be in
one of 16 topological SC phases.  Those 16 classes of SC phases have
different even/odd numbers of electrons on even-by-even, even-by-odd,
odd-by-even, and odd-by-odd lattices.  This result can be easily
generalized to any dimensions.  We find that there are 256
($2^{(2^d)}$) different topological SC orders in 3-dimensions
($d$-dimensions).  Such topological SC orders are robust against weak
interactions that do not break the translation symmetry.

This research is supported by NSF Grant No. DMR-0706078, NFSC no. 10228408,
and NFSC no. 10874017.


\end{document}